\documentclass[aps, prx, superscriptaddress, twocolumn, longbibliography, tightenlines, showpacs, floatfix, pdftex]{revtex4-1}

\usepackage{amsmath}
\usepackage{amssymb}
\usepackage{bm}
\usepackage{bbm}
\usepackage{float}
\usepackage{tikz} 
\usepackage{graphicx}
\usepackage{amsmath}
\usepackage{amssymb}

\def\id{\mathbbm{1}}
\def\integers{\mathbb{Z}}
\newcommand{\corr}[1]{\langle #1\rangle}

\begin{document}
\title{Generalized liquid crystals:  giant fluctuations and the vestigial chiral order of $I$, $O$ and $T$ matter}
\date{\today}

\author{Ke Liu}
\affiliation{Instituut-Lorentz for Theoretical Physics, Universiteit Leiden,
PO Box 9506, NL-2300 RA Leiden, The Netherlands}
\author{Jaakko Nissinen} 
\affiliation{Instituut-Lorentz for Theoretical Physics, Universiteit Leiden,
PO Box 9506, NL-2300 RA Leiden, The Netherlands}
\author{Robert-Jan Slager}
\affiliation{Instituut-Lorentz for Theoretical Physics, Universiteit Leiden,
PO Box 9506, NL-2300 RA Leiden, The Netherlands}
\author{Kai Wu}
\affiliation{Stanford Institute for Materials and Energy Sciences, SLAC National Accelerator Laboratory and Stanford University,
Menlo Park, California 94025, USA}
\author{Jan Zaanen}
\affiliation{Instituut-Lorentz for Theoretical Physics, Universiteit Leiden,
PO Box 9506, NL-2300 RA Leiden, The Netherlands}

\begin{abstract}
{The physics of nematic liquid crystals has been subject of intensive research since the late 19th century. However, the focus of this pursuit has been centered around uni- and biaxial nematics associated with constituents bearing a $D_{\infty h}$ or $D_{2h}$ symmetry respectively. In view of general symmetries, however, these are singularly special since nematic order can in principle involve any point group symmetry. Given the progress in tailoring nano particles with particular shapes and interactions, this vast family of ``generalized nematics" might become accessible in the laboratory. Little is known since the order parameter theories associated with the highly symmetric point groups are remarkably complicated, involving tensor order parameters of high rank.  Here we show that the generic features of the statistical physics of such systems can be studied in a highly flexible and efficient fashion using a mathematical tool borrowed from high energy physics: discrete non-Abelian gauge theory. Explicitly, we construct a family of lattice gauge models encapsulating nematic ordering of general three dimensional point group symmetries. We find that the most symmetrical ``generalized nematics" are subjected to thermal fluctuations of unprecedented severity.  As a result, novel forms of fluctuation phenomena become possible. In particular, we demonstrate that a vestigial phase carrying no more than chiral order becomes ubiquitous departing from high point group symmetry chiral building blocks, such as  $I$, $O$ and $T$ symmetric matter.}
\end{abstract}


\maketitle

\section{Introduction}
\begin{figure}[h!]
\center
\includegraphics[width=0.4\textwidth]{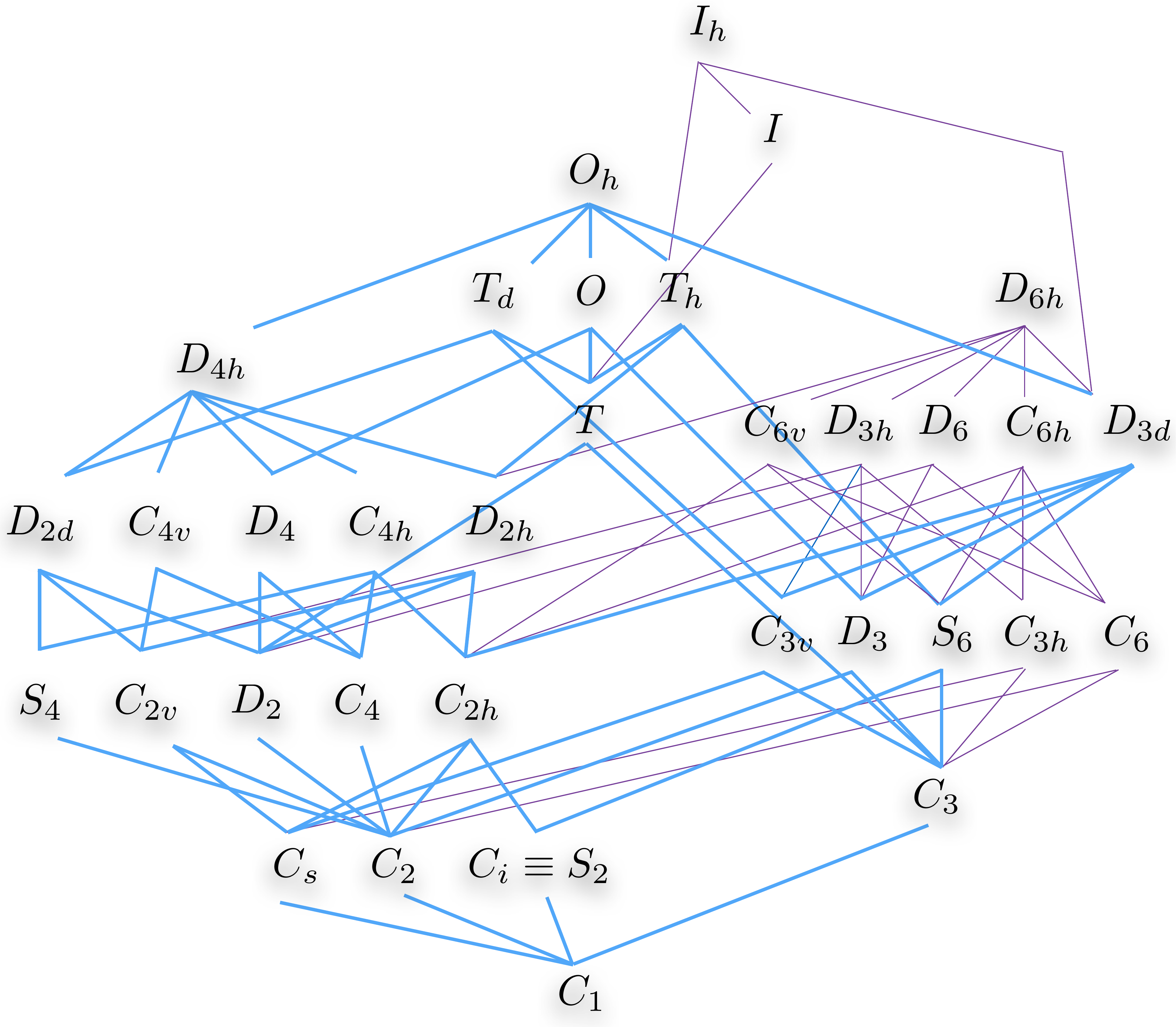}
\caption{
A selection of the rich hierarchy of three-dimensional point groups (subgroups of $O(3)$). In the context of generalized nematics, each point group can correspond to a liquid crystalline phase, giving rise to many more nematics besides the conventional uniaxial ($D_{\infty h}$) and biaxial ($D_{2h}$) cases. The lines indicate subgroup relations (not unique) and spontaneous symmetry breaking may occur between any two groups connected by them. Here the $O_h$ case is highlighted as an example, showing that one can descend down all the way to $C_1$ corresponding to full symmetry breaking; all nematic phases related to the symmetries connected by the thick lines are allowed.}
\label{fig:subgroups}
\end{figure}

The subject of ``vestigial" or ``mesophase"  (intermediate temperature) order was born in the theater of classical molecular matter in the form of nematic, cholesteric and smectic liquid crystals \cite{Friedel22, DeGennesProst95}.  Although this is a very mature field, it has been suffering from the limitation that mesophases are experimentally only easily formed departing from ``rod-like" degrees of freedom, giving rise to uniaxial nematics \cite{DeGennesProst95, Onsager49, MaierSaupe59, LebwohlLasher72}. In addition, a substantial literature is devoted to biaxial nematics with $D_{2h}$ symmetry, associated with plate-like constituents or
``mesogens" \cite{YuSaupe80, SeveringSaalwachter04, AcharyaPrimakKumar04, MadsenEtAl04, BiaxialBook, Freiser70, Alben73, Straley74, BiscariniZannoni1995, Fel1991}.
However, departing from the Landau-de Gennes symmetry paradigm, these represent only two examples of a vast family of potential phases: in principle matter can break the $O(3)$ rotational symmetry of three-dimensional ($3D$) space down to any of its subgroups, i.e., 3D point groups, for our conventions see \cite{SternbergBook, Michel2001}.
Furthermore, the hierarchy of $O(3)$ subgroups leaves room for rich sequences of vestigial phases.
We show an example of this hierarchy by a selection of $3D$ point groups in Fig. \ref{fig:subgroups}.  Dealing with, say, $C_2$ degrees of freedom or ``mesogens", a cascade of phases like $O_h \rightarrow D_{4h} \rightarrow C_{4v} \rightarrow C_{2v} \rightarrow C_2$ could be realized in principle  upon lowering temperature, pending the right microscopic interactions between the constituents. Chemistry has been proven to be an intricate affair in this regard, but new opportunities open up with the advances in the manufacturing of nanoparticles and colloids \cite{Matijevic1981, LiJosephsonStein2011, Mark2013, HuangEtAl15} that can be given particular shapes, while there is potentially quite a bit of control over their mutual interactions \cite{Glotzer2007, DamascenoGlotzer2012, vanAndersGlotzer2014, Manoharan2015, HuangEtAl15}. 

As a further impetus for this pursuit, we construct a general framework in terms of a lattice model that can incorporate all three dimensional point group symmetries and therefore is ideal to study generalized nematics. Concretely, we achieve this by a generic lattice formulation of discrete non-Abelian gauge theory. The model allows us to expose some spectacular, generic traits of the statistical physics arising for the most symmetric point groups. 
  We demonstrate that their order parameters are subjected to thermal fluctuations of unprecedented intensity.  This is illustrated in Fig. \ref{Fig:phase_diagram} where we show the transition temperatures ($T_c$'s) of the various nematics relative to a common reference --- their mean field temperatures would be identical. 
The actual $T_c$'s of the highly symmetric $I$, $O$ and $T$ are nonetheless extremely reduced by thermal fluctuations, as compared to the transition temperatures of less symmetric nematics. In addition, we uncover that {\em chiral} symmetry plays a very special role. This associates to the breaking of $O(3)$ to its subgroup of proper rotations $SO(3)$ and refers to the familiar property that molecules can be left- or right handed, as in e.g. sugar water formed exclusively from left-handed glucose molecules, leading to well-known optical activity. However, we now depart from chiral point groups describing mixtures of equal number of left- and right handed species which are subjected to {\em spontaneous} symmetry breaking, resulting in the formation of  {\em chiral nematics} and {\em chiral liquids}, reminiscent of the recently pointed out domains of well-defined broken chirality in liquid phases of systems of $C_{2v}$ symmetric ``bent-core" mesogens \cite{HoughEtAl2009} and  chiral liquids melted from cubic crystalline phases \cite{Dressel2014}.
Moreover, we show that, even under the  most adverse conditions for vestigial order to occur, a {\em chiral liquid} still splits the isotropic liquid from the full nematic order for the most symmetric $I,O$ and $T$ chiral point groups, as consequence of the extreme thermal fluctuations of the full orientational order. 

Why has this spectacular statistical physics not been discovered a long time ago? After all, constructing the theory of three-dimensional orientational order should be a well-defined exercise in the Landau paradigm of spontaneous symmetry breaking. However, the Landau-de Gennes order parameter theory of more symmetric nematics generically involves a complicated high-rank tensor order parameter theory \cite{NelsonToner81, AllenderLeeHafiz85, GramsbergenLongadeJeu1986, Jaric86, Fel1995, Fel1995b, RadzihovskyLubensky2001, LubenskyRadzihovsky2002, Mettout2006, AllenderLonga08}, making the physical ramifications are basically unexplored, in spite of the identification of the general structure of point group invariants \cite{JaricSharp1984, Michel2001}. In this sense the problem represents one of the remaining frontiers of the Landau paradigm. Indeed, dealing with more complicated point groups, one has to generalize the familiar uniaxial order parameter: $\mathbb{Q}_{ab} = \frac{3}{2}n_a n_b - \frac{1}{2} \delta_{ab}$, in terms of a vector $\vec{n} = (n_x, n_y, n_z)$, to complicated higher rank tensors (up to rank 6 for $I_h$ the symmetry of a regular icosahedron), cf.  Appendix \ref{SimandOrderparameters}.
Notwithstanding, we find that a mathematical edifice borrowed from high-energy physics is remarkably efficient in computing universal and generic features of the associated statistical physics: 3D $O(3)/G$ lattice gauge theory, where the gauge group $G$ describes the discrete point groups associated with the nematic ordering. Although such gauge theoretical setups have been also considered in the context of spin liquids with unaxial and biaxial symmetries \cite{Grover07, Grover11, XuLudwig2012}, these ideas in the present context are rooted in the seminal observation of Refs. \cite{LammertRoksharToner93, LammertRoksharToner95}  that a particularly simple $O(3)$-vector $\integers_2$ gauge incarnation encodes for the uniaxial nematic order parameter in 3D. Accordingly, this already led to a recent extension encapsulating the full family of 2D nematic orders  in terms of the (Abelian) 2D point groups ($SO(2)/\integers_{p}$) \cite{LiuEtAl15}. From a theoretical perspective, the surprises of the statistical physics of generalized nematics are thus a manifestation of the richness of discrete gauge theories involving the in general non-Abelian three dimensional point groups. 

The rest of this paper is organized as follows. In Section \ref{SecII}, we review the $O(3)$-vector $\mathbb{Z}_{2}$ lattice gauge theoretical formulation of uniaxial nematics and introduce a generalized non-Abelian lattice gauge theory formulation that is able to cope with nematics of arbitrary point group symmetries, including the familiar $D_{\infty h}$-uniaxial and $D_{2h}$-biaxial cases.
In Section \ref{sec:results} we discuss various nematic transitions in three dimensions and expose the severity of the orientational fluctuations in case of highly symmetric nematic phases, highlighting the practical advantages of our gauge formulation with regards to numerical simulations. Moreover,  we address the emergence of a fluctuation induced chiral  liquid. Finally, in Sec. \ref{sec:Conclusions}, we conclude with an outlook of associated experimental realizations and further applications of our gauge theory.

\section{Generalized nematic phases and gauge theoretical formulation}\label{SecII}

The symmetry breaking framework of orientational order is straightforward to address in the context of the subgroups structure in Fig. \ref{fig:subgroups}. The associated physics of the nematic phases can then be studied in terms of Landau-de Gennes theory, where an order parameter tensor is needed for each subgroup of $O(3)$ \cite{NelsonToner81, AllenderLeeHafiz85, GramsbergenLongadeJeu1986, Jaric86, Fel1995, RadzihovskyLubensky2001, LubenskyRadzihovsky2002, Mettout2006, AllenderLonga08}. Instead of the Landau-de Gennes free energy, we can consider a lattice model for the coarse grained order parameter tensors. The lattice model should offer a realization of the phase transition(s) associated with the Landau classification \cite{MaierSaupe59, LebwohlLasher72, YuSaupe80, Freiser70, Alben73, Straley74, BiscariniZannoni1995, Romano06, Romano08, BiaxialBook, Nissinen16}. However, the construction of the order parameters is a non-trivial problem in itself, and there is the additional task of enumerating the parameters in the free energy or lattice model that classify the phases. In most cases, these goals have been achieved only to a degree by improvising in specific simplified cases and the resulting generic classification of three-dimensional nematic phases remains therefore quite unexplored.

It goes therefore without mentioning that a uniform framework to explore this rich landscape of generalized nematics in a {\it systemic} fashion would be a value asset to the active research fields concerned with generalized nematic order. This should have also direct bearing on the experimental side in the long term. Indeed, although it has been pointed out a long time ago that nematics phases formed out of "platelets", i.e. mesogens having $D_{2h}$ symmetry, can in principle give rise to generalized biaxial nematics \cite{Freiser70}, only recently the stabilization has been quantified in terms of anisotropy in the
constituents  and interactions \cite{SeveringSaalwachter04, AcharyaPrimakKumar04, Merkel04}, see also \cite{Tschierske10} for a review. Furthermore, the $C_6 (N+6)$ phase of DNA comprises an experimentally observed example of a nematic phase having another symmetry than the familiar uniaxial one \cite{Toner83, Podgornik96}.  Finally, to the best of our knowledge, the only other specific mesogenic systems that have received considerable attention are those carrying $C_{2v}$ symmetry. These 'banana' shaped constituents have most importantly also been studied in the context of experiment \cite{ Takezoe06} as well as in theoretical setups \cite{LubenskyRadzihovsky2002}. We do point out that in these instances the mesogens appear to organize into more complicated aggregates in the observed liquid crystal, columnar and smectic phases. Nonetheless, they motivate the relevance of the pursuit of generalized nematics that are captured within our comprehensive gauge theoretical description.

As already observed in Refs. \cite{LammertRoksharToner93, LammertRoksharToner95}, see also \cite{Mettout2006}, the uniaxial nematic point group symmetry can be incorporated as gauge symmetry on coarse grained local degrees of freedom of a lattice model, instead of the director order parameter tensor $\mathbb{Q}_{ab}$. Moreover, the gauge symmetries give rise to an explicit way of incorporating the topological defects in to the model and an effective way to generate the order parameters \cite{Nissinen16}. We here generalize this approach to all three dimensional point groups. However, before turning to the problem of nematics with general point group symmetries that highlight the intricacies of the non-Abelian nature, let us first review the gauge theoretical description for uniaxial nematics.

\subsection{Uniaxial nematics and $\mathbb{Z}_2$ gauge theory} \label{section:Z2}
The $D_{\infty h}$-uniaxial order can be captured by an $O(3)$-vector model coupled to a $\integers_2$ gauge theory, turning the order parameter vector $\vec{n}$ into a director (the rod) with a head-to-tail symmetry. The simplicity in the gauge formulation is rooted in the  Abelian $\mathbb{Z}_2$ nature of the uniaxial $D_{\infty h}$ symmetry acting on $\vec{n}$.
More specifically, to  describe the coarse grained order parameter theory, one departs from an auxiliary cubic lattice regulating the short-distance cut-off of the theory. The theory has variables $\sigma^{z}_{ij}=\pm 1$ living on the bonds $\corr{ij}$ of the lattice, that interact by a plaquette term $-K \sum_{^{l}_{i}\Box_{j}^{k}} \sigma_{ij}^z\sigma^{z}_{jk}\sigma^z_{kl}\sigma^{z}_{li} $ thereby defining Wegner's Ising gauge theory \cite{Kogut79}. To describe nematics, the gauge fields are minimally coupled to nearest-neighbor $O(3)$ vectors $\vec{n}_i$ on the sites of the lattice via a Higgs term $-J \sum_{\langle ij \rangle} \sigma^z_{ij} \vec{n}_i \cdot \vec{n}_j$.

Despite its simplicity, the Ising lattice gauge theory is actually enough to elucidate the nature of non-perturbative discrete gauge theories in general \cite{FradkinShenker79}. For large $J$ the matter and gauge fields are ordered via the Higgs mechanism. The coupling $K$ controls the gauge fields and for small $K$,$J$ the gauge fields are confined, effectively ``gluing" the matter fields to gauge invariant singlets not unlike quark confinement in hadrons. For large $K$ and small $J$, matter is disordered while the gauge fields are ``ordered" forming a deconfining phase with topological gauge fluxes as excitations. Although realizations of  such ``topological nematic phases" \cite{LammertRoksharToner95} have been identified in strongly interacting electron systems \cite{SenthilFisher2000, SenthilFisher2001, NussinovZaanen2002, PodolskyDemler2005}, deconfinement seems unphysical dealing with  ``molecular" matter. Therefore the regime of interest is strong gauge coupling $K \rightarrow 0$, where one finds the fully ordered ``Higgs phase" and a fully disordered confining phase. These encode for the uniaxial nematic phase and the isotropic liquid, respectively. The  gauge symmetry identifies $\vec{n}_i \simeq - \vec{n}_i$ and as a result, the physical gauge invariant observables correspond with the {\em directors} $\mathbb{Q}_{ab}$. 
Consequently, upon integrating out the fluctuating gauge fields at $K \rightarrow 0$,  one obtains an effective theory of the de-Gennes kind \cite{LebwohlLasher72}: $H \sim \sum_{\langle i,j\rangle}  \mathrm{Tr}~ \mathbb{Q}_{i} \cdot \mathbb{Q}_{j}$, with the $\mathbb{Q}_{i}$ being the uniaxial tensor order parameter \cite{DeGennesProst95}. 

\begin{figure}
\center
\includegraphics[width=0.45\textwidth]{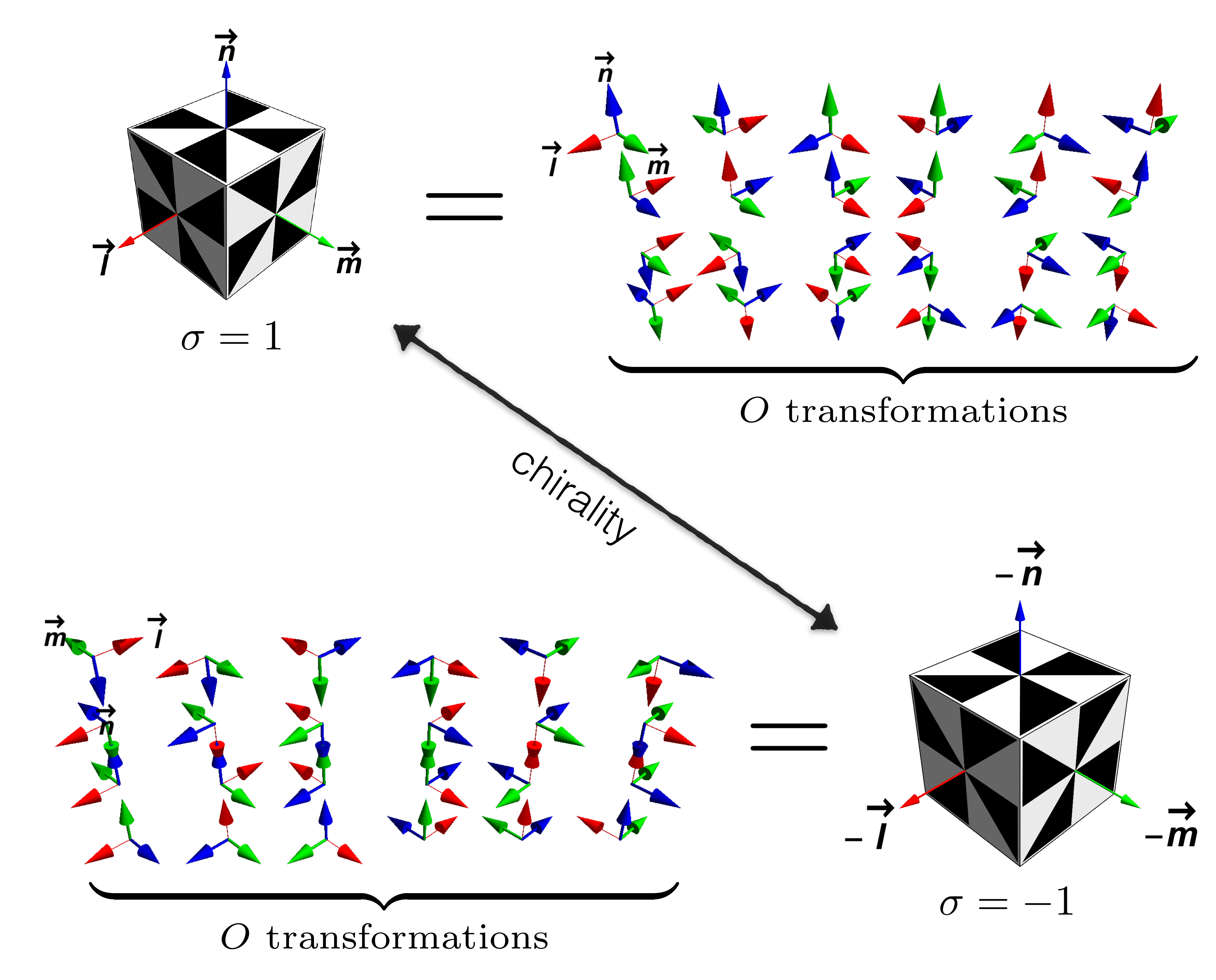}
\caption{The correspondence between the shapes of ``mesogens'' and the gauge symmetries acting on the rotors (triads). Here $O$-symmetric "mesogens" (cubes) are considered as an example  \cite{AshcroftMerminBook}. Top: The orientations of the "mesogen" correspond to triads $R_i$. Acting with $O$ gauge transformations on the triads shows orientations that leave the mesogens invariant, and are realized as gauge symmetries in the model Eq.\eqref{eq:gaugetheory}. Bottom: When the symmetry group of the mesogen is a proper point group, there are two mesogens that have the same symmetry but opposite chirality. Chirality is indicated by the black-white ``propeller'' scheme of the cube or left- and right-handed triads. Imposing the chirality $\sigma\to-\sigma$ as a symmetry leads to $O_h$ symmetry. }
\label{fig:gaugecartoon}
\end{figure}

\subsection{Generalized nematics and non-Abelian gauge theory}
For the uniaxial case only a single ``nematic" vector along the $D_{\infty h}$ axis is required and the $\integers_{2}$ gauge symmetry then simply turns the vector into the director. However, in case of a general 3D nematic ``mesogen", one has to introduce a discrete lattice gauge theory that can cope with any of the non-abelian point group symmetries $G\subset O(3)$. Accordingly, one should depart from 3D matrix rotor fields $R_i$ defined in terms of an orthonormal triad as 
\begin{equation}
R_i=(\begin{matrix} \vec{l}_i & \vec{m}_i & \vec{n}_i \end{matrix})^T \in O(3),
\end{equation}
instead of a \emph{single} vector. The $O(3)$ constraints make the vectors $\vec{n}_i^{\alpha} = \{\vec{l}_i,\vec{m}_i,\vec{n}_i\}_{\alpha=1,2,3}$ to an orthonormal triad that is right or left handed pending on 
\begin{equation}
\sigma_i = \det(R_i)= \vec{l}_i \cdot (\vec{m}_i \times \vec{n})_i = \pm 1.
\end{equation}
These chiral degrees of freedom will be discussed in detail in Sec. \ref{sec:Chiral}. The generalized nematic symmetry implies identifications of the triad as 
\begin{align}
R_i\simeq \Lambda_i R_i, \quad \vec{n}_i^{\alpha} \simeq \Lambda_i^{\alpha\beta}\vec{n}^{\beta},\quad \forall\Lambda_i\in G,
\end{align}
generalizing the $\integers_2$ director symmetry of the \emph{single} vector $\vec{n}_i$.  In very concrete terms, the gauge symmetries on the $O(3)$-triad $R_i$ thus encode for the physical degrees of freedoms of ``mesogens" with the symmetry $G$ in their ``body-fixed" frame. See Fig. \ref{fig:gaugecartoon} for an illustration.
On the other hand, global rotations $\Omega\in O(3)$ defined by a space-fixed frame and distinguished from the local body-fixed rotations of the mesogens, act on the triads as 
\begin{align}
R_i \to R_i \Omega^{\mathrm{T}},\quad \vec{n}_i^{\alpha} \to \Omega \cdot \vec{n}_i^{\alpha}, \quad \Omega \in O(3). \label{eq:globalO3}
\end{align}
where the $\cdot$ denotes ordinary matrix multiplication of the vector $\vec{n}_i^{\alpha}$.

The so-called $O(3)/G$ lattice gauge theory can be formulated by introducing such degrees on an auxiliary cubic lattice with $O(3)$ triads $R_i$ defined on sites and gauge fields $U_{ij} \in G$ defined on links,
\begin{align}
\label{eq:gaugetheory}
\beta H &= -\sum_{\langle ij \rangle} \textrm{Tr}~\big[R_i^T \mathbb{J} U_{ij} R_j\big] 
- \sum_{\square} \sum_{\mathcal{C}_\mu} K_{\mathcal{C}_\mu} \delta_{\mathcal{C}_{\mu}}(U_{\square}) \mathrm{Tr}\big[U_{\square}\big],
\end{align} 
where the invariance of a point-group-symmetric ``mesogen'', which leads to the identification $R_i \simeq \Lambda_i R_i$, is realized by the gauge transformation
\begin{align}
R_i \to \Lambda_i R_i, \quad U_{ij} \to \Lambda_i U_{ij} \Lambda_j^T, \quad \forall \Lambda_i \in G.
\end{align}
In addition the model has the global $O(3)$-rotation symmetry Eq. \eqref{eq:globalO3}. 

The first term in Eq.\eqref{eq:gaugetheory} models the orientational interaction between $G$-symmetric ``mesogens'', where 
$\mathbb{J}$ is a symmetric coupling matrix encoding the nematic ``exchange" terms, which is invariant under $G$:  $\Lambda \mathbb{J} \Lambda^{\mathrm{T}} = \mathbb{J}$, $ \forall \Lambda \in G$. 
In standard gauge theory language, this term is nothing but a Higgs term \cite{FradkinShenker79} for the matter fields $R_i$. 
In the current context, however, the central importance lies in the fact that it favors alignment of $G$-symmetric ``mesogens'' and thus can realize spontaneous symmetry breaking from an isotropic $O(3)$ liquid phase to a nematic phase having point group symmetry.

The second term in Eq.\eqref{eq:gaugetheory} is a defect suppression term, generalizing the $K$ term of the $\mathbb{Z}_{2}$ case. It involves oriented products of gauge fields $U_{\square} = \prod_{\corr{ij} \in \square} U_{ij}$ around plaquettes $\square$ of the lattice. Plaquettes with non-zero gauge flux or field strength, $U_{\square} \neq \id$, represent topological defects in the nematic. Under a gauge transformation $\Lambda_i$, $U_\square \to \Lambda_i U_\square \Lambda_i^{-1}$ and therefore is defined only up to conjugation. Correspondingly, $K_{\mathcal{C}_{\mu}}$ denotes the core energy of the defects with the flux $U_{\square}\in G$ and is a function of the conjugacy classes $\mathcal{C}_{\mu}$ of the group $G$ since defects in the same conjugacy class are physically equivalent. These gauge defects do not directly classify topological defects in nematics, but they are closely related via the so-called Volterra construction \cite{Friedel64, Kleinert89b}.
The nematic defects are usually classified by homotopy groups of the manifold $O(3)/G$ \cite{Mermin1979, Michel1980} which is the order parameter space of the $G$-nematic and as well the low-energy manifold of the model Eq.\eqref{eq:gaugetheory} in the Higgs phase. Disordered configurations in the Higgs term can be suppressed by assigning a finite core-energy to the gauge defects.
Thus, $K_{\mathcal{C}_{\mu}}$ can effectively be regarded as tuning the fugacity of the nematic defects. As we are however interested in the ordinary nematic to isotropic transitions, we will not consider the gauge field dynamics associated with finite $K_{\mathcal{C}_{\mu}}$ and set these parameters to zero in the remainder.

\subsection{Gauge theory and nematic phases}
All physics in the model Eq. \eqref{eq:gaugetheory} follows from gauge invariant quantities, as by Elitzur's theorem \cite{Elitzur75} correlation functions of gauge non-invariant quantities vanish. 
This is in direct analogy to the $\integers_2$ or $D_{\infty h}$-case, where one describes the physics in terms of gauge invariant tensor which is director order parameter $\mathbb{Q}_{ab}$.  In the general case, the relevant order parameters are high-rank tensors that are linear combinations of tensor constructed from the rotors $(R_i)^{\alpha}_a = (\vec{n}_i^{\alpha})_a$. Formally, they can be  expressed as
\begin{align}
\mathbb{O}_i^{(G)} &= \sum_{ \{\lambda \}} c_\lambda \mathbb{O}_i^{\lambda},  \nonumber \\
(\mathbb{O}^{\lambda=\alpha\beta\cdots \gamma}_{i})_{a b\cdots c} &= (R_i)^{\alpha}_a(R_i)^{\beta}_b \cdots(R_i)^{\gamma}_c,
\end{align}
where $\lambda = (\alpha,\beta,\dots, \gamma)$ is a multi-index. The above tensors transform under the gauge group $G$ on the indices $\alpha$ and as vectors under global $O(3)$-rotations on the indices $a$. The order parameters are obtained as the averages of $\corr{\mathbb{O}^{(G)}}$ and these tensors are specified by their rank and tensor-symmetries. We note that the tensors $\mathbb{O}_i^{\lambda}$ are not all independent due to the $O(3)$ constraints on the $R_i$. 

The gauge theory realizes $G$-nematic ordering by guaranteeing that when the $O(3)$ symmetry spontaneously breaks, $ \langle \mathbb{O}^{G}_{i} \rangle \neq 0$, but all non gauge-invariant combinations of $(R_i)^{\alpha}_a$ vanish. One observes that the theory Eq. \eqref{eq:Jeff} can in fact act as an order parameter generator, since gauge-invariant quantities can be constructed via e.g. integrating out gauge fields. 
This is one of the advantages of the gauge theoretical description over traditional methods such as 
Landau-de Gennes theories and lattice modelss for nematic ordering, as these methods rely the on the relevant order parameters tensors as input  \cite{MaierSaupe59, LebwohlLasher72, Freiser70, Alben73, Straley74, Romano06, Romano08, LongaPajakWydro09, BiaxialBook}. 
Moreover, though these traditional methods have been proven to be very fruitful for nematics with relatively simple symmetries such as $D_{\infty h}$ and $D_{2h}$, they are quite involved for general point groups.
In this regard, one may consider the $I_h$-icosahedral nematic, whose order parameter is a rank-$6$ traceless tensor of the form (Appendix \ref{appendix:T_O_I_orderparameters}):
\begin{align}\label{eq:Ih_orderparameter_main} 
\mathbb({\mathbb{O}}^{I_h}_{i})_{abcdef} \!=&  \frac{112}{5} \! \sum_{\text{cyclic}} \!
\big[ 
\vec{l}_{i}^{\otimes 6}  \! + \! \sum_{ \{+,- \}} \!
\big(
\frac{1}{2} \vec{l}_{i} \! \pm \! \frac{\tau}{2} \vec{m}_{i} \! \pm \! \frac{1}{2\tau} \vec{n}_{i}
\big)^{\otimes 6}
\big]_{abcdef} \nonumber \\
& -\frac{16}{5} \sum_{\text{permutations}} \delta_{ab} \delta_{cd} \delta_{ef},
\end{align}
where $^{\otimes n}$ denotes the tensor power, ``cyclic" refers to all cyclic permutations of $\{ \vec{l}, \vec{m}, \vec{n} \}$, $\sum_{\{+,-\}}$ sums over all four combinations of the two $\pm$ signs, $\tau = \frac{1+\sqrt{5}}{2}$ is the golden ratio, and ``permutations" runs over all non-equivalent combinations of indices. 
An order parameter lattice model of the form $ H \sim -\sum_{\corr{ij}} \mathrm{Tr} \ \mathbb{O}^{I_h}_i\cdot \mathbb{O}^{I_h}_j$ can be obtained from the gauge theory Eq. \eqref{eq:gaugetheory} by integrating out the gauge fields in a high- or low-temperature expansion, generalizing the lattice models of the uni- and biaxial nematics \cite{Nissinen16}. 
However, needless to say, $\mathbb{O}^{I_h}_{i}$ contains an abundant number of terms making the corresponding order parameter theory inevitably complicated.

On the other hand, the gauge theory Eq. \eqref{eq:gaugetheory} is convenient for nematics of arbitrary point group symmetries.
It requires the symmetry of nematics only as input for fixing degrees of freedom of the gauge fields $U_{ij}$ and fits all point groups in a universal framework.
Therefore, it is a remarkably efficient device to study generalized nematic ordering.

Finally, returning to the subgroup structure in Fig. \ref{fig:subgroups}, we remark that, within the gauge model, a subset of intermediate phases in Fig. \ref{fig:subgroups} can be realized by simply tunning temperature and the coupling matrix $\mathbb{J}$. 
These involve generalized biaxial-uniaxial-liquid transitions for axial groups 
$\{C_n, C_{nv}, S_n, C_{nh}, D_n, D_{nh}, D_{nd} \}$ where an anisotropic $\mathbb{J}$ is possible. 
We will discuss in a separate work the anisotropic couplings in Eq. \eqref{eq:gaugetheory} in order to bring the anisotropy-induced intermediate phases of the $O(3)$ subgroup hierarchy into play \cite{LiuEtAl2016b}. 

\begin{figure*}
\centering
\includegraphics[width=0.85\textwidth]{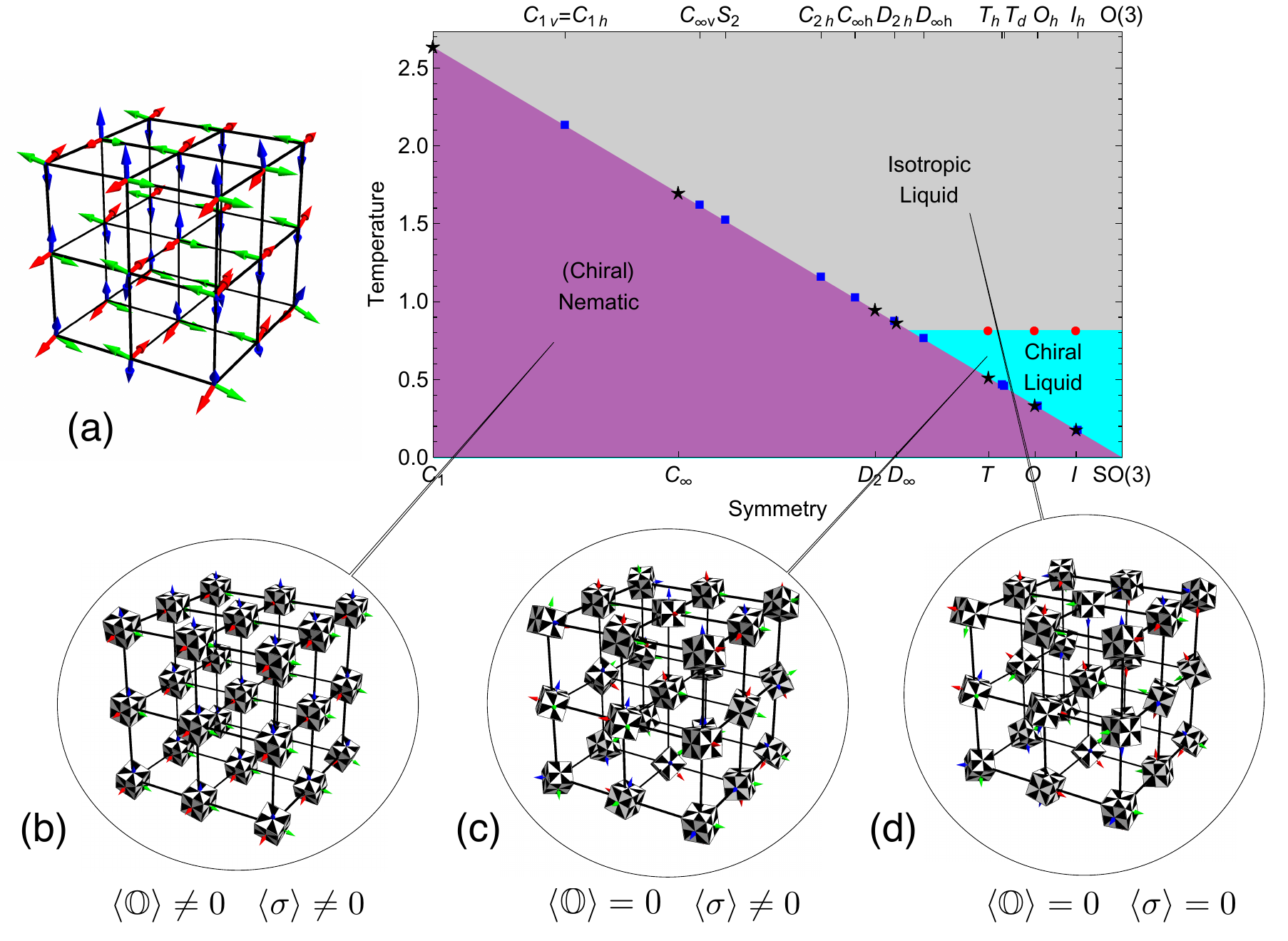}
\caption{ 
Phase diagram of different point-group-symmetric nematics. The vertical axis is the temperature in units of $J$, the bottom and top axis label proper and improper point groups, respectively, in the order of increasing symmetry. 
The nematic transition temperature, shown by stars for point groups and squares for improper point groups, is decreasing with increasing symmetry due to thermal fluctuations in the orientational order. These fluctuations become huge for highly symmetric groups, and lead to the emergence of a vestigial chiral liquid phase for the $I$, $O$ and $T$ nematics.
The transition temperature, shown by filled circles, of the chiral liquid phase to the fully disordered isotropic liquid phase is identical in the gauge theory Eq.\eqref{eq:gaugetheory} because of a common orientationally disordered background. 
The results presented here are studied on a cubic lattice(\textbf{a}), where the triad matter fields reside on sites and the gauge fields are defined on links (black), and with isotropic coupling $\mathbbm{J} = J\id$.
The long range ordering associated with the three phases is shown in the insets by using the $O$-nematic as an example [Fig. 2]:
 (\textbf{b}) nematic which is full ordered; (\textbf{c})  chiral liquid which has no orientation order but has a preferred chirality;  and (\textbf{d}) isotropic liquid which is fully disordered. Note that cubes in (\textbf{c}) have  a preferred black-and-white color scheme, while in (\textbf{d}) the two color schemes appear randomly. For details of the simulations, see Appendix \ref{SimandOrderparameters}.}
 \label{Fig:phase_diagram}
\end{figure*}

\section{Results and simulations} \label{sec:results}
As outlined in the previous section, the gauge theory Eq.\eqref{eq:gaugetheory} is an efficient and flexible framework for generalized nematics, and it is straightforward to simulate numerically using the standard Metropolis Monte-Carlo algorithm. 
As mentioned, in order to focus on nematic-isotropic liquid (Higgs-confinement) transitions, we have set $K_{\mathcal{C}} = 0$, i.e. the defects do not have an explicit core energy.
Phase transitions can then be easily detected by monitoring the strength of the generalized nematic ordering, $q = \sqrt{\mathrm{Tr}[\langle \mathbb{O}^{\boldsymbol{\lambda}}_{ab...c} \rangle]^2}$ and the specific heat  (Appendix \ref{appendix:ordering_strength}).
We accordingly simulated all three dimensional crystallographic point groups, the icosahedral groups $\{I,I_h\}$ and the five infinite axial point groups $ \{C_{\infty}$, $C_{\infty v}$, $C_{\infty h}$, $D_{\infty}$, $D_{\infty h}$\} .

\subsection{Giant thermal fluctuations of highly symmetric nematics}\label{section:fluctuations}
The results for a large number of representatives are collected in Fig. \ref{Fig:phase_diagram}, where the vertical axis is the reduced temperature and the horizontal axes arbitrarily accommodate point groups in increasing order of symmetry. 
In Fig. \ref{Fig:phase_diagram} the isotropic coupling $\mathbb{J} = J \id$ has been chosen for simplicity.
A remarkable observation here is the huge thermal fluctuations for nematics of highly symmetric point groups as evidenced by the extremely low transition temperatures. The trivial $C_1$ nematic sitting in the bottom of Fig.\ref{fig:subgroups} has the highest transition temperature $T_c$ which consistently decreases towards $T=0$ as one ascends the subgroup hierachy towards $O(3)$. This is surprising because with the isotropic coupling matrix $\mathbb{J}=J\id$, a naive mean field theory would predict all nematics to have the same $T_c$, whereas thermal fluctuations in three dimensional systems typically reduce $T_c$ by a modest $\sim 20\%$ \cite{GerberFisher74}.
However, dealing with the most symmetric icosahedral $\{ I, I_h \}$ nematics, this reduction is more than an order of magnitude! One thus immediately notices the symmetry hierarchy in Fig.\ref{fig:subgroups}.

To understand the physics better, let us first zoom in on the $C_1$ ``nematic" having the highest transition temperature. 
This incarnates triads having no symmetry and describes a non-linear $O(3)$ matrix model maximally breaking the rotational symmetry \cite{NelsonToner81}.
Moving to $\{ C_{\infty}, C_{\infty v} \}$, $\{ D_2, D_{2h}\}$ and $\{ D_{\infty}, D_{\infty h} \}$ cases, the geometric interpretation of the mesogens become cones, cuboids and cylinders, respectively. 
Climbing further up the hierarchy of Fig. \ref{fig:subgroups}, the triads turn into tetrahedrons $\{T, T_d, T_h \}$, cubes $\{O, O_h \}$ and  icosahedra $\{I, I_h \}$.
It is intuitively clear that towards spheres $\{SO(3), O(3) \}$ sitting in the top of the symmetry hierarchy, the differences between the ordered state and isotropic space are increasingly harder to discern and the thermal fluctuations associated with the order will increase in severity. 

We emphasize that in Fig. \ref{Fig:phase_diagram} the isotropic coupling $\mathbb{J}$ is taken, so that thermal fluctuations are roughly equal for all three axes defined by the triads. This is important for the axial point groups $\{C_n, C_{nv}, S_n, C_{nh}, D_n, D_{nh}, D_{nd} \}$, whose geometric interpretation as mesogens is in terms of (colored) $n$-gonal prisms \cite{AshcroftMerminBook}. These nematics are characterized by a primary order parameter for the main axis and a secondary order parameter in the perpendicular plane.
When $n$ increases, the $n$-gonal prisms become more cone- or cylinder-like, and the in-plane fluctuations hence are more severe and tend to restore the in-plane $O(2)$ symmetry, while ordering is easy along the main axis just as in the $n=\infty$ case. 
Thus we cannot simply incorporate these cases into Fig. \ref{Fig:phase_diagram}, since to properly quantify the influence of these in-plane fluctuations, we need coupling matrices $\mathbb{J}$ with anisotropic entries. However, we can already conclude that the same trend is also true for in-plane fluctuations of the axial nematics. Consequently, the remarkable power of the gauge theory Eq. \eqref{eq:gaugetheory} allows for a common microscopic reference for all different point groups, making it always possible to compare the orientational fluctuations in absolute terms.

Finally, though in Fig. \ref{Fig:phase_diagram} the gauge theory Eq. \eqref{eq:gaugetheory} has been studied in the $K_{\mathcal{C}} = 0$ limit, we have preliminarily checked by our simulations that until sufficiently large $K_{\mathcal{C}}$, the results remain qualitatively similar to those in the $K_{\mathcal{C}} = 0$ limit. Therefore the features that have been discussed are stable against finite $K_{\mathcal{C}}$.
A large $K_{\mathcal{C}} $ will suppress the defects and therefore the disordering, leading to the phase transitions moving to higher temperatures. 
For large enough $K_{\mathcal{C}}$ and small enough Higgs couplings, the theory will feature a confinement-deconfinement phase transition of the gauge fields \cite{Kogut79}. With deconfined gauge fields and disordered rotors, the physics is quite different and entails a regime of (non-Abelian) topological excitations and topological order. Although such deconfinement phenomena have been identified in strongly correlated electron systems \cite{SenthilFisher2000, SenthilFisher2001, NussinovZaanen2002, PodolskyDemler2005}, there are no identified analogues in thermal liquid crystal systems.

\begin{figure}
\includegraphics[width=0.45\textwidth]{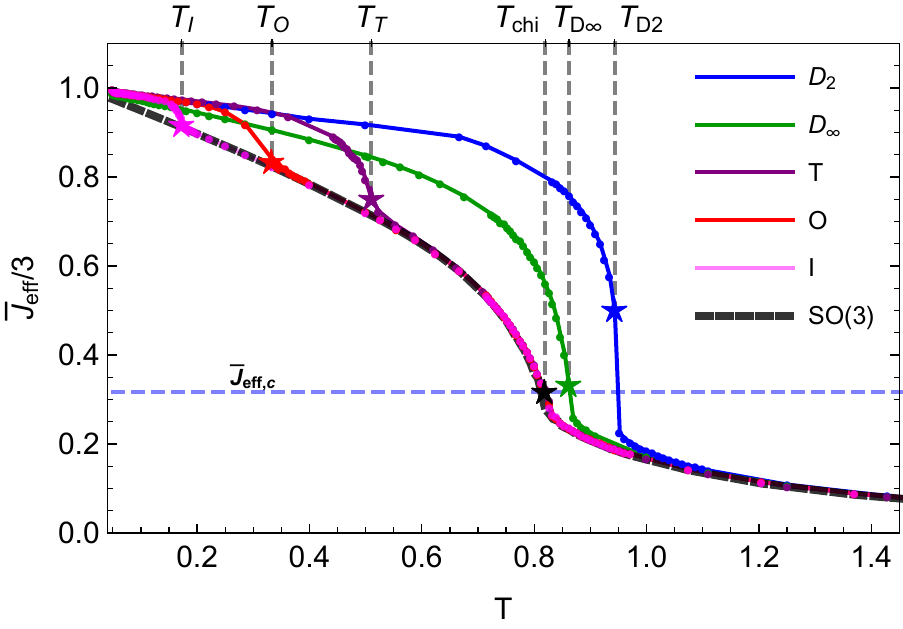}  
\centering
\caption{ The mean-field coupling, $\overline{J}_{\rm eff} $ defined by Eq. \eqref{eq:Jeff}, measuring the short range nematic correlations, as a function of the temperature (in units of $J$). Transition temperatures of the chiral and nematic transitions are indicated by stars and the group $SO(3)$ (black) is shown as a reference. The dashed blue line shows a critical mean-field coupling $\overline{J}_{\textrm{eff},c} \simeq 0.95$ where the chiral order sets in.
Due to huge fluctuations in the nematic order, $\overline{J}_{\rm eff}$ exceeds the critical value at the chiral transition at $T_{\rm  chi} \simeq 0.82$ for the highly symmetric groups $I$, $O$ and $T$, which is well above the transition to the nematic phase. In contrast, for the less symmetric $D_{\infty}$ and $D_2$ groups, $\overline{J}_{\rm eff}$ only exceeds the critical value at the nematic transition by a sudden jump.}
\label{fig:Jeff}
\end{figure}

 \subsection{The chiral liquid as a fluctuation driven vestigial phase}\label{sec:Chiral}

Another remarkable result of the gauge theory Eq. \eqref{eq:gaugetheory} at $K_{\mathcal{C}} = 0$ (and small but finite $K_{\mathcal{C}}$) limit is the emergence of a chiral liquid phase (the light blue region in Fig. \ref{Fig:phase_diagram}) for $T$, $O$ and $I$ nematics.
This phase \emph{spontaneously} breaks the $O(3)$ symmetry of the isotropic liquid to $SO(3)$ and is characterized by chiral order and the absence of any orientational order.
From the symmetry point of view, this phase is in principle possible for all proper point groups $\{C_n, D_n, T, O, I \}$, i.e., subgroups of $SO(3)$.
However, according to our simulations, it \emph{only} occurs as ``vestigial" phase for highly symmetric $T$, $O$ and $I$ nematics before the nematic full order sets in. Furthermore, the chiral transition temperatures are identical within the accuracy of our simulations.

The mechanism at work for the chiral liquid is an elegant mixture of intricacies of the point group symmetries and fluctuation physics: the chiral phase arises in essence via order-out-of-disorder, as we now demonstrate.

The chiral symmetry is related to the {\em central (or commuting) group} of inversions $\integers_2 = \{\id,-\id \}$ in $O(3) = SO(3)\times \integers_2$. Due to the Abelian nature of the inversions, the pseudoscalar chirality 
$\sigma_i = \det R_i  = \pm 1$ can be always factored out. The rotors can be parametrized with composite fields as $R_i = \sigma_i \widetilde{R}_i$, where $\widetilde{R}_i \in SO(3)$ is a rotation matrix of pseudo-vectors. 
In the context of our gauge model, the fields  $\{\sigma_i\}$ are well-defined and physical (gauge-invariant) for gauge groups composed of only proper rotations. We note that the chemical potential for the difference $\sum_i \sigma_i$ of right- and left-handed ``mesogens" is formally equivalent to a non-zero ``magnetic field" $h\neq 0$ for the Ising variables $\{\sigma_i\}$. In the following, we will always set this chemical potential to zero meaning that the chirality $\sigma_i$ is a locally fluctuating variable with no bias for the sense of handedness. The Hamiltonian Eq. \eqref{eq:gaugetheory} at $K_{\mathcal{C}} = 0$ thus can be rewritten as
\begin{align} 
\beta H_{K_{\mathcal{C}} = 0} = -\sum_{\langle ij\rangle} \sigma_i \sigma_j \textrm{Tr} \big( \widetilde{R}^T_i \mathbb{J} U_{ij} \widetilde{R}_{j} \big), \label{Eq:sigma-out}
\end{align} 
featuring explicit global $O(3) =SO(3)\times  \integers_2 $ symmetry that can be broken separately.

One notices that the $SO(3)$ part, $\widetilde{q}_{ij}=\mathrm{Tr } [\widetilde{R}_i^T U_{ij}\widetilde{R}_j]$, in the Hamiltonian Eq. \eqref{Eq:sigma-out} provides an effective coupling for the chiral Ising fields and vice versa.
$\widetilde{q}_{ij}$ measures the short range correlations of the orientational order. 
As discussed in Subsection \ref{section:fluctuations}, depending on the underlying symmetry, fluctuations in the orientational order can be relatively mild or extremely severe.
In contrast, the Ising order $\sigma_i$ in three dimensions is subjected to rather benign fluctuations.
Hence, when fluctuations in the orientational order are severe, the effective short-range coupling for the chiral Ising fields induced by $\widetilde{q}_{ij}$ can be very strong before the full nematic order sets in, and cause the chiral Ising fields to order.

To verify the the validity of the above scenario, we can take a mean-field approximation by defining an average Ising coupling $\overline{J}_{\rm eff}$ of short-range correlations as
\begin{equation} \label{eq:Jeff}
\overline{J}_{\rm eff} = \frac{1}{N_{\rm bonds}} \sum_{\corr{ij}}\langle \mathrm{Tr} ( \tilde{R}^T_{i} \mathbb{J} U_{ij} \tilde{R}_j) \rangle,
\end{equation}
and measure it in our simulation.
In Fig.  \ref{fig:Jeff} we show $\overline{J}_{\rm eff}$ for various point group symmetries. 
As expected, in the case of highly symmetric $T, O,I$ nematics, $\overline{J}_{\rm eff}$ builds up smoothly and becomes strong enough for the chiral Ising order well above the nematic transition temperature,
whereas for the less symmetric cases (e.g. $D_2$, $D_{\infty}$) $\overline{J}_{\rm eff}$ is small before the chiral order sets in, then changes abruptly when a direct isotropic-nematic transition taking place (see Appendix \ref{appendix:chiral} for more details).

Fig. \ref{fig:Jeff} also reveals a peculiarity of the ``microscopic physics" hard wired in the gauge model, that should surely not be taken literally dealing with physical nematic systems. One observes that at temperatures well above the nematic transitions, the $\overline{J}_{\mathrm{eff}}$'s for the various point groups coincide. This includes the ``baseline" of the $SO(3)$ point group featuring only the chiral phase. However, rather than just having a temperature independent chiral coupling, there is still quite some action going on in the orientational sector. The remaining orientational fields $\widetilde{q}_{ij}$ are then described by $SO(3)/SO(3)$ gauge theory, albeit coupled to the chiral degrees of freedom. The temperature dependence is set by a famous gauge theory contraption \cite{FradkinShenker79}: when the matter fields are in the fundamental representation of the gauge group, for small $K_{\mathcal{C}}$ the theory is described by weakly interacting fields on the links and the Higgs phase (at large $J$) becomes indistinguishable from confinement (small $J$). The details are of interest to gauge theorists, and what matters in the present context is that it adds a temperature dependence of independent $SO(3)$ bond-fields to the effective chiral coupling which has no relationship to the microscopic physics of the condensed matter system. For a mean-field calculation of the chiral transition for the Ising-$SO(3)/SO(3)$ theory, see Appendix \ref{appendix:chiral}.

Nevertheless, again the benefit of the gauge formulation is a common reference frame to compare the fluctuations in absolute terms. One infers from Fig.  \ref{fig:Jeff} that only rather close to the nematic transitions, when the ordering sets in, the $\overline{J}_{\mathrm{eff}}$'s ``peel off" the $SO(3)/SO(3)$ reference line. The tendency towards chiral order is hard wired in the gauge model to be the same for all chiral point groups but is controlled by the orientational fluctuations. The extra correlations associated with the full point group symmetry are of importance only close to the full ordering. In the $I,O$ and $T$ cases this happens at much lower temperatures than the chiral transition and therefore their chiral transition temperatures are very nearly identical (within the accuracy of our simulations). However, referring to the same ``common gauge", the intrinsic fluctuations of the less symmetric nematics (e.g. $D_{\infty}, D_2$) are just too weak to disorder the orientational fields and leave room for the chiral vestigial phase.

The mechanism discussed above has actually quite a history in the context of the magnetism of iron-based superconductors \cite{FernandesChubukovSchmalian14,KamiyaKawashimaBatista11}, featuring ``stripe antiferromagnets" breaking not only internal spin symmetry, but also spatial rotational symmetry. Departing from a square lattice ($C_4$ symmetry) in two dimensions, the $x$ and $y$ directions become inequivalent (``nematic" $C_2$ symmetry) in the striped antiferromagnet, involving an Ising-type symmetry breaking. Generically one finds that either first the ``nematic" order sets in followed by the 
full magnetic order at a somewhat lower temperature, or both occur in a single merged first order transition \cite{FernandesChubukovSchmalian14}. 
A qualitatively similar mechanism is invoked to explain these observations (e.g., see Refs. \cite{KamiyaKawashimaBatista11, FernandesChubukovSchmalian14}), with the lattice symmetry taking the role of chiral symmetry with the antiferromagnet replacing the $O(3)$-rotations breaking phase. However, in terms of a classical unfrustrated spin model featuring the symmetries of the striped antiferromaget, it is impossible to stabilize the vestigial ``$C_2$-nematic" phase in three dimensions \cite{KamiyaKawashimaBatista11}. This is because the thermal fluctuations of the classical $O(3)$ spins are falling short in this regard as compared to highly symmetric nematics and one has to resort to microscopic frustration physics of the iron-based materials to boost the thermal fluctuations. We note that despite these symmetry considerations, the precise quantum-mechanical mechanism of the $C_2$ nematic phase in this complicated system with itinerant and localized physics is still subject of considerable debate \cite{FernandesChubukovSchmalian14, Beak2015, Chubukov2015,Liang2013}.
  
\section{Conclusions and outlook}\label{sec:Conclusions}

There is a rich landscape of ``generalized nematics", formed and fully classified in terms of the 3D point group 
symmetries. Still waiting to be fully explored, they represent a remaining frontier of the Ginzburg-Landau order parameter paradigm involving order parameters of unprecedented complexity. 

In this paper, we have introduced a lattice gauge theory model that realizes generalized nematic ordering in three dimensions and incorporates all point groups. We further mobilized the machinery of discrete non-Abelian gauge theory, discovering that it is remarkably powerful in addressing {\em generic} features of the statistical 
physics of such systems. In addition to the generalized nematics phases, we identified  a vestigial \emph{chiral} liquid phase that arises for nematics systems with high point group symmetries. 
This chiral liquid phase is associated with a locally fluctuating handedness of the mesogens and emerges via spontaneous symmetry breaking at intermediate temperatures between the nematic phase and a normal liquid phase, possessing only short-range orientational order but long-range chiral order.
Based on a mean-field calculation and the transition temperature observed in our simulations, we find that the isotropic-chiral transition is in the Ising universality class with a in principle diverging correlation length at the transition.
Nevertheless, beyond the mean-field treatment, fluctuations in the orientational order may cause the transition to become (weakly) first order, even though the short-range orientational order develops smoothly during this transition.
Accordingly, finite size scaling analysis needs to be performed in order to verify the nature of the chiral-isotropic phase transition and the associated physical quantities in our model.

Interestingly, in a recent experimental work by Dressel et al. \cite{Dressel2014} on studying phenyl-thiophene-based polycatenar compounds, a chiral liquid phase melting from a cubic crystalline phase was in fact observed before a transition to an isotropic liquid. This intermediate liquid phase exhibited domains of opposite chirality when analyzed under a polarizer, whereas peaks were observed in differential-scanning-calorimetry measurements in the transitions to the cubic crystalline phase and the isotropic liquid. Most importantly, they identified that the crystalline order was short-ranged in the chiral liquid phase and acted as the main agent stabilizing the chiral phase. Moreover,  the short-range crystalline order increased continuously in the isotropic-chiral transition. These observation are consistent with our predictions of the order-out-of-disorder mechanism with the caveat that in their system, the cubic symmetric phase is crystalline and our orientational model does not by construction include the translational ordering. Their experimental methods are suited to our context as well, where the orientational anisotropy of the generalized nematic phase is in principle observable in X-ray diffraction or bifringence experiments \cite{Toner83}.

In general the gauge model has somewhat of a status of a ``spherical cow" compared to the intricacies facing the experimentalists in the soft matter laboratories.
Nevertheless, when it comes to isolating the physical principle at work under a single framework, one can also view it as the limit of Platonic perfection. As stated above, 
to realize these in the laboratory, one needs preferably building blocks with $T, O$ or $I$ symmetry. As a case in point of the strange traits of chemistry, such molecules
are {\em extremely rare} and we have only found a few examples for each. An example of a $T$-symmetric molecule is the $[\text{Ga}_4\text{L}_6]^{12-}$ tetrahedral metal-ligand cluster \cite{JohnsonRaymond01}. Very recently nanosized ``giant" tetrahedra have been fabricated by placing different polyhedral oligomeric silsesquioxane (POSS) molecular nanoparticles at the vertices of a rigid tetrahedral framework. These tetrahedra have in addition tunable hydrophilic interactions and have been observed in self-assembled crystalline and alloyed supramolecular quasicrystalline phases induced by entropic packing \cite{HuangEtAl15}. The even more exuberant $O$ representative 
is the well-studied transporting protein Ferritine that stores and releases iron in organisms \cite{HonarmandBillHagedoornHagen01}. 
Finally,  the chiral icosahedral symmetry
is found in the form of viruses \cite{ Zandietal04} including the common rhinovirus \cite{KumarBlaas13}. We note, however, that all these cases in fact
involve very complex molecules in the nano-scale.
Nevertheless, it appears that there is a realistic potential to overcome the experimental challenges of the control over the shapes and interactions of nano-particles and colloids. We hope that our theoretical insights might act as a source of inspiration for the experimental community to build systems with such intricate spatial symmetries and find out whether for instance the vestigial chiral order can be realized in the laboratory. 

Viewed from a fundamental theoretical perspective, the non-Abelian gauge theories associated with point groups are highly interesting by themselves, and arriving at a timely moment. We have only 
explored a small corner (large gauge coupling, minimal extra structure) of the full portfolio of these theories.  Dealing with the Higgs (generalized nematic) phases there is interesting work to 
do, such as further exploring the nature of the topological defects occurring in the high symmetry point groups \cite{KlemanFriedel08}. Moreover, 
upon increasing the gauge coupling $K_{\mathcal{C}}$ a landscape of {\em deconfining} phases will appear \cite{Kogut79, FradkinShenker79} that remains to be charted. This has its merit in yet a quite 
different field of physics. Deconfining states of discrete gauge theories play a crucial role in the subject of {\em topological order} and topological quantum computation \cite{MathyBais07, Kitaev2006, Wen1991, AndristMartin-Delgado2011, XuLudwig2012}, often limited to two-spatial dimensions or Abelian symmetries. The point group symmetries form natural building blocks to extent this to the non-Abelian realms in three dimensions.

\begin{acknowledgments}
We would like to thank Z. Nussinov for related collaborations and E. Cobanera, H. W. J. Bl\"{o}te, D. Kraft and B. van Zuiden for helpful discussions. 
This work has been supported by the Dutch
Foundation on the Research of Fundamental Matter (FOM), which is part of NWO. K. L. is supported by the State Scholarship Fund program organized by China Scholarship Council (CSC). 
K. W. is supported by DOE-BES Division of Materials Sciences and Engineering DMSE at Standford University.
\end{acknowledgments}

\begin{appendix}

\section{Details of simulations and generalized nematic order parameters}\label{SimandOrderparameters}

The phase diagram in Fig. \ref{Fig:phase_diagram} was determined by simulating the lattice gauge theory Eq. \eqref{eq:gaugetheory} by the Metropolis Monte-Carlo on lattices of sizes $L^3 = 8^3,\dots, 24^3$. To ensure the thermalization of our ensembles, we monitored the results for both cooling from a random initial state as well as heating from a uniform ordered initial state.

Using our gauge theory formulation, we have calculated the associated nematic tensor order parameters, as discussed above, and the the pseudo-scalar chiral order parameter $\sigma_i = \vec{l}_i \cdot (\vec{m}_i \times \vec{n}_i)$. Phases in the phase diagram are defined as
\begin{align} \label{define phases}
  \begin{cases}
               \langle \mathbb{O}^{G}_{ab\dots c} \rangle \neq 0 \quad & \text{nematic}\\
               \langle \mathbb{O}^{G}_{ab\dots c} \rangle = 0, \langle \sigma \rangle \neq 0 \quad & \text{chiral} \\
               \langle \mathbb{O}^{G}_{ab\dots c} \rangle = 0, \langle \sigma \rangle = 0 \quad & \text{isotropic},
            \end{cases} 
\end{align}
where $\mathbb{O}^{G}$ is the order parameter tensor with $G$ rotational symmetry, $\corr{\cdots}$ denotes the thermal average, $\mathbb{O} = L^{-3} \sum_i \mathbb{O}_i$ and $\sigma = L^{-3} \sum_{i} \sigma_i$.

\subsection{The $T,O, I$ tensor order parameters} \label{appendix:T_O_I_orderparameters}
We will present here the order parameters for $T$-, $O$- and $I$-nematics, which realize the chiral liquid phase. 

\emph{$T$-invariant order parameter.}
The tetrahedral-$T$ group contains $12$ proper rotations leaving a tetrahedron invariant.  It can be defined by the following set of generators,
\begin{align}
\hat{c}_2(\vec{n}) = \left(
\begin{array}{ccc}
 -1 & 0 & 0 \\
 0 & -1 & 0 \\
 0 & 0 & 1
\end{array}
\right),  \
\hat{c}_3(\vec{l}+\vec{m}+\vec{n})= \left(
\begin{array}{ccc}
 0 & 1 & 0 \\
 0 & 0 & 1 \\
 1 & 0 & 0
\end{array}
\right),
\end{align}
where $\hat{c}_2(\vec{n})$ is a two fold rotation of the body axis $\vec{n}$,
and $\hat{c}_3(\vec{l}+\vec{m}+\vec{n})$ is a three fold rotation about the axis $\vec{l}+\vec{m}+\vec{n}$ which generates cyclic permutations of the three body axis,
\begin{align} \label{T generators}
\hat{c}_2(\vec{n}) R_i = \big[-\vec{l} \ -\vec{m} \ \vec{n} \big]^{\mathrm{T}}_i, \ 
\hat{c}_3(\vec{l}+\vec{m}+\vec{n})R_i = \big[\vec{m} \quad \vec{n} \quad \vec{l}\big]^{\mathrm{T}}_i. 
\end{align}
Here and in the following we will use the notation $\hat{c}_p (\vec{a})$ to denote a $p$-fold rotation about an axis $\vec{a}$.

The local order parameter of a $T$-nematic defined  by only ``matter'' fields needs to be invariant under Eq.\eqref{T generators} and all its combinations, which is realized by a rank-$3$ tensor of the form 
\begin{align} \label{T order paramameter}
  \mathbb{O}^{T}_{ abc, i} =   \sum_{\text{cyclic}} 
 \big ( \vec{l} \otimes \vec{m} \otimes \vec{n} \big)_{abc,i},
\end{align}
where the sum $\sum_{\text{cyclic}}$ runs over cyclic permutations of $\{\vec{l}, \vec{m}, \vec{n} \}$.

\emph{$O$-invariant order parameter.}
The octahedral-$O$ group consists of all $24$ proper rotations leaving a cube invariant. A set of generators is given by $\{ \hat{c}_4(\vec{n}) , \hat{c}_3(\vec{l}+\vec{m}+\vec{n}), \hat{c}_2(\vec{m}+\vec{n})\}$,
where $\hat{c}_3(\vec{l}+\vec{m}+\vec{n})$ is same as in Eq.\eqref{T generators}, and $\hat{c}_4 (\vec{n})$ and $\hat{c}_2(\vec{m}+\vec{n})$ are given as 
\begin{align} \label{O generators}
\hat{c}_4(\vec{n}) = \left(
\begin{array}{ccc}
 0 & -1 & 0 \\
 1 & 0 & 0 \\
 0 & 0 & 1
\end{array}
\right), \ 
\hat{c}_2(\vec{m}+\vec{n}) = \left(
\begin{array}{ccc}
 -1 & 0 & 0 \\
 0 & 0 & 1 \\
 0 & 1 & 0
\end{array}
\right).
\end{align}

The simplest $O$-invariant order parameter is a rank-$4$ tensor, $\mathbb{O}^{O_h}_{abcd,i}$, and the chiral field $\sigma_i$. The $\mathbb{O}^{O_h}_{abcd,i}$ is an order parameter for a $O_h$ nematic, with the form
\begin{align} \label{Oh order paramameter}
\mathbb{O}^{O_h}_{abcd,i} = & \frac{5}{2} \big(\vec{l}^{\otimes 4} + \vec{m}^{\otimes 4} + \vec{n}^{\otimes 4} \big)_{abcd,i} 
-\frac{1}{2} \big( \delta_{ab}\delta_{cd} + \delta_{ac}\delta_{bd}  \nonumber \\
& + \delta_{ad}\delta_{bc} \big),
\end{align}
where $^{\otimes n}$ denotes tensor powers, by which $l^{\otimes 4}_i = \vec{l}_i\otimes \vec{l}_i \otimes \vec{l}_i\otimes \vec{l}_i$.
Eq.\eqref{Oh order paramameter} is $O_h$ invariant, in addition to the invariance of Eq.\eqref{O generators}, it is also invariant under spatial inversion.
However, a nontrivial order in $\sigma_i$ will leave the set $\mathbb{O}^O_i = \{\mathbb{O}^{O_h}_{abcd, i}, \sigma_i \}$ only $O$ invariant.
In principle, we can also define a tensor invariant only under $O$, but it requires a rank of $5$ with more complexity.

\emph{$I$-invariant order parameter.}
The icosahedral group $I$ consists of all $60$ proper rotations that leave a icosahedron invariant. An icosahedron centered at $(0, 0, 0)$ is given by its $12$ vertexes at \cite{Litvin91}
\begin{align}
(\pm \frac{1}{2}, \  0, \  \pm \frac{\tau}{2}),
(\pm \frac{\tau}{2},\  \pm \frac{1}{2},\  0), 
(0, \ \pm \frac{\tau}{2},\  \pm \frac{1}{2}),
\end{align} 
where $\tau = (\sqrt{5}+1)/2$ is the golden ratio. It is invariant under a five fold rotations about its six diagonals. The axis $\vec{l}+\tau\vec{n}$ is the diagonal passing trough vertices
$(-\frac{1}{2}, 0,  -\frac{\tau}{2})$ and $(\frac{1}{2}, 0, \frac{\tau}{2})$. A set of generators is given by $\{ \hat{c}_5 (\vec{l}+\tau \vec{n}), \hat{c}_3(\vec{l}+\vec{m}+\vec{n}), \hat{c}_2(\vec{n}) \}$, 
where $\hat{c}_3(\vec{l}+\vec{m}+\vec{n})$ and $ \hat{c}_2(\vec{n})$ are the same as those in Eq.\eqref{T generators}, $\hat{c}_5 (\vec{l}+\tau \vec{n})$ is given by 
\begin{align} \label{I generators}
\hat{c}_5(\vec{l}+\tau\vec{n}) = \left(
\begin{array}{ccc}
 1/2 & -\tau/2 & 1/(2\tau) \\
 \tau/2 & 1/(2\tau) & -1/2 \\
 1/(2\tau) & 1/2 & \tau/2
\end{array}
\right).
\end{align}
Similar to the $O$-nematic case, an $I$-invariant order parameter consists of an orientational part and a chiral part,
$\mathbb{O}^I_i = \{ \mathbb{O}^{I_h}_{abcdef,i}, \sigma_i\}$. 
The orientational part, $\mathbb{O}^{I_h}_{abcdef,i}$, is an order parameter for the $I_h$-nematic,
\begin{align} \label{Ih order paramameter}
\mathbb{O}_{abcdef,i}^{I_h} \!=&  \frac{112}{5} \! \sum_{\text{cyclic}} \!
\big[ 
\vec{l}^{\otimes 6}  \! + \! \sum_{ \{+,- \}} \!
\big(
\frac{1}{2} \vec{l} \! \pm \! \frac{\tau}{2} \vec{m} \! \pm \! \frac{1}{2\tau} \vec{n}
\big)^{\otimes 6}
\big]_{abcdef,i} \nonumber \\
& -\frac{16}{5} \sum_{\text{permutations}} \delta_{ab} \delta_{cd} \delta_{ef},
\end{align}
where ``cyclic" means all cyclic permutations of $\{ \vec{l}, \vec{m}, \vec{n} \}$, $\sum_{\{+,-\}}$ sums over all four combinations of the two $\pm$ signs and ``permutations" runs over all non-equivalent combinations of indices, making $15$ delta functions terms in total. A purely $I$-invariant single tensor also exists, but it is at least of rank-$7$.

\subsection{Strength of the nematic ordering}\label{appendix:ordering_strength}
The nematic interaction in Eq. \eqref{eq:gaugetheory} prefers alignment of triads so that the components of the nematic order parameter with $G$ rotational symmetry $\mathbb{O}^{G}_{abc...,i}$ develop an expectation value. This will lead to the two point correlation function behaving as (repeated indices are summed over)
\begin{align}
\lim_{ \lvert i-j \rvert \rightarrow \infty} 
\langle \mathbb{O}^G_{abc..., i} \mathbb{O}^G_{abc..., j} \rangle = 
		\begin{cases}
				\ \langle \mathbb{O}^G_{abc...} \rangle^2 >0, \ &\text{nematic} \\
				\ 0, \ &\text{otherwise}. 
		\end{cases} 
\end{align}
The order parameter tensor has $G$ rotational symmetry. This allows us to define a strength of the nematic ordering as
$q = \sqrt{\langle \mathbb{O}^G_{ab\dots c} \rangle^2}$,
by which phases defined in Eq.\eqref{define phases} can be equivalently defined as 
\begin{align} \label{re-define phases}
  \begin{cases}
               q \neq 0, \ \sigma \neq 0 \quad & \text{nematic}\\
               q = 0, \  \sigma  \neq 0 \quad & \text{chiral} \\
               q = 0, \ \sigma  = 0 \quad & \text{isotropic},
            \end{cases} 
\end{align} 
where $\sigma = \langle \sigma_i \rangle$ measures the strength of the long range chiral order.

The phase transitions can be located by the peak of the susceptibility of the nematic order and the chiral order, 
$\chi(q)$ and $\chi(\sigma)$, defined as
\begin{align}
\chi(q) &=  \frac{L^3}{T} \big(\corr{q^2} - \corr{q}^2 \big), \\
\chi(\sigma) &=  \frac{L^3}{T} \big(\corr{\sigma^2} - \corr{\sigma}^2 \big), 
\end{align}
As summarized in Fig.\ref{Fig:phase_diagram}, for the $T$-, $O$- and $I$-nematic $\chi(q)$ and $\chi(\sigma)$ peak at different temperatures indicating two phase transitions respect to the nematic order and the chiral order, while for others $\chi(q)$ and $\chi(\sigma)$ peaks coincide.

To corroborate of the strength of the nematic ordering $q$ as a probe of the phase transition, we also compute the heat capacity defined as  
\begin{equation}
c_v = \frac{1}{T^2L^3} \big(\corr{E^2} - \corr{E}^2 \big),
\end{equation}
where $E$ is the internal energy.
Consistent with the results by computing $\chi(q)$ and $\chi(\sigma)$,
$c_v$ exhibits two well separated peaks for the $T$-, $O$- and $I$-nematics coincided with the peak of $\chi(q)$ and $\chi(\sigma)$, while one peak for others which do not support the chiral phase.

\section{Detailed analysis of the fluctuation induced chiral phase}\label{appendix:chiral}
 
\begin{figure}
\centering
\includegraphics[width=0.45\textwidth]{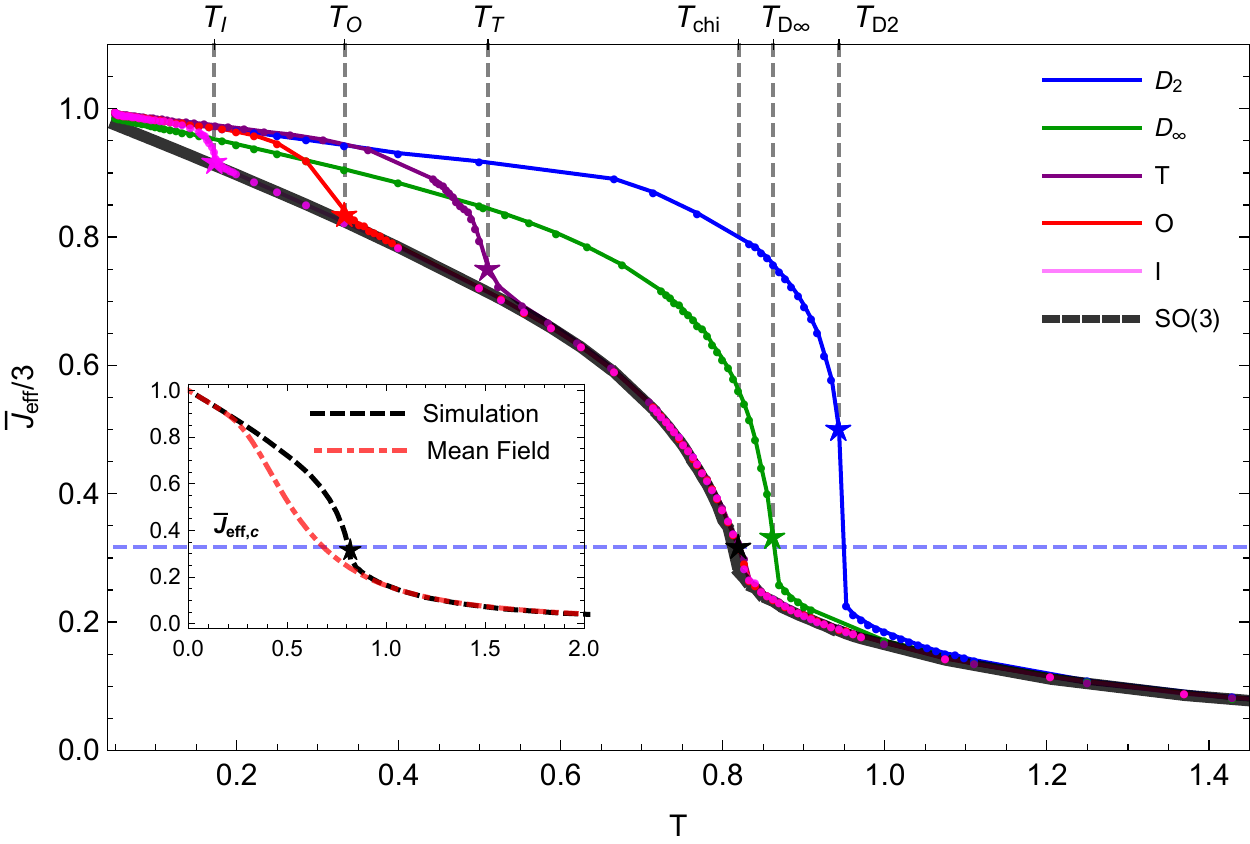}  
\caption{ Comparison of  $\overline{J}_{\rm eff}$ for $G=SO(3)$ case computed by the mean-field approximation Eq. \eqref{eq:JeffMF_SO3} and by Monte Carlo simulations. Data in Fig. \ref{fig:Jeff} are reproduced here for convenience.}
\label{fig:JeffMF_SO3}
\end{figure}

With the mean field approximation Eq. \eqref{eq:Jeff}, the Hamiltonian Eq. \eqref{Eq:sigma-out} can be rewritten as
\begin{align} 
\beta H_{K_{\mathcal{C}} = 0} &= -\sum_{\corr{ij}} J_{\mathrm{eff}, ij}(\beta) \sigma_i \sigma_j, \nonumber \\
 J_{\mathrm{eff}, ij} &= \beta \overline{J}_{\mathrm{eff}} + \delta J_{\mathrm{eff},ij}.
\end{align} 
$J_{\textrm{eff}, ij}(\beta)$ can be viewed as an effective coupling for the Ising fields, and can be compute analytically under a further mean field approximation (Appendix \ref{sec:SO3calc}) in the degenerate limit of gauge group $G=SO(3)$. In this case the matter-gauge variables are independent fields $W_{ij} = \widetilde{R}_i U_{ij} \widetilde{R}^T_j \in SO(3)$ interacting only via the coupling to the chiral Ising variables. By performing a partial integration over the $SO(3)$ fields we determine $J_{\textrm{eff},ij}(\beta)$ as a power series, and by solving $J_{\textrm{eff}, ij}(\beta_{\rm chi}) = \beta_{c, \rm 3D Ising}$ to high order we obtain the chiral temperature $\beta_{\rm chi} \simeq 1.2064$ in perfect agreement with our Monte-Carlo value. Similarly, $\overline{J}_{\rm eff}(\beta)$ can be calculated in a mean-field approximation for the Ising fields, and the curve $\overline{J}_{\rm eff,MF}$ is shown in Fig. \ref{fig:JeffMF_SO3} (inset) in comparison to the $G=SO(3)$ Monte-Carlo data. As expected the mean-field result $\overline{J}_{\rm eff, MF}$ is correct in the high and low temperature regions. We clearly see the correlations of the chiral Ising and orientational degrees of freedom amplifying the $SO(3)$ ordering very close to the chiral transition where the curves start to deviate. One also observes from Fig. \ref{fig:JeffMF_SO3} that below the chiral transition the system indeed finally realizes that the orientational symmetries are actually different with the effect that the transition temperatures to the full nematic order are quite different. This is to be expected based on the different fluctuations of the nematic order parameters, as discussed above.

\subsection{Calculations for $G=SO(3)$}\label{sec:SO3calc}

In order to quantify the interdependence of the orientational and Ising degrees of freedom, we can compute the coupling $J_{\rm eff}(\beta)$, as well as $\overline{J}_{\rm{eff}}$ in a mean-field approximation for the Ising fields in the case of  gauge group $G=SO(3)$. There is no phase transition in the orientational degrees of freedom \cite{FradkinShenker79} and the effective action with the coupling $J_{\mathrm{eff}, ij}(\beta)$ for the Ising fields is always well-defined. Specifically, the partition function takes the form
 \begin{align*}
 Z_{SO(3)} &=  \sum_{\{\sigma_i\}} \int \mathcal{D}\{\widetilde{R}_i, U_{ij}\} \prod_{\corr{ij}} e^{\beta \mathrm{Tr} [\widetilde{R}_i U_{ij} \widetilde{R}_{j}] \sigma_i\sigma_j} \nonumber \\
 &=  \sum_{\{\sigma_i\}} \prod_{\corr{ij}} \int_{SO(3)} \mathrm{d} W_{ij}~ e^{\beta \mathrm{Tr} [W_{ij}] \sigma_i\sigma_j} \nonumber \\
 &= \sum _{\{\sigma_i\}} e^{-\sum_{\corr{ij}} J_{\mathrm{eff},ij}(\beta) \sigma_i\sigma_j},
 \end{align*}
where in the second line, due to the $SO(3)$ gauge symmetry, we can always pick a gauge where $\widetilde{R}_{i} = \id$ for all $i$ and a new $SO(3)$ link variable $W_{ij} = \widetilde{R}_i U_{ij} \widetilde{R}^T_j$ and trivially integrate over the matter fields $\{\widetilde{R}_i\}$ and $U_{ij} \in SO(3)$ with the Haar measure $\frac{1}{8\pi^2}\int_{SO(3)} \mathrm{d} g$ = 1, see \cite{FrankelBook}. We determine $J_{\rm{eff}}(\beta)$ from the odd-power series expansion of $\log f(x)$ with 
\begin{align}
f(x) &= \frac{1}{8\pi^2}\int_{SO(3)} \mathrm{d} W e^{x \mathrm{Tr}[W]} \nonumber\\
&= \frac{1}{8\pi^2}\int_{S^2} \mathrm{d}\Omega \int^{\pi}_0 \mathrm{d} \varphi~  2(1-\cos \varphi) e^{x(1+2\cos \varphi)} \nonumber \\
&=e^{x} \big( I_0(2 x)-I_1(2 x) \big), 
\end{align}
where $S^2$ is the two-sphere with volume element $\mathrm{d} \Omega$ and $I_n(x)$ is the modified Bessel function of the first kind. Here we used that $\mathrm{Tr}[W(\widehat{n},\varphi)] = 1+2 \cos \varphi$, where $\varphi\in [0,\pi)$ is the angle of rotation and $\widehat{n}$ the axis of rotation of the element $W\in SO(3)$. The angle $\varphi$ also determines the conjugacy classes of $SO(3) \simeq \mathbb{RP}^3$. The measure $\mathrm{d} \Omega \mathrm{d} \varphi (1-\cos \varphi)$ satisfies $\mathrm{d} g = \mathrm{d}(h g) =\mathrm{d}(g h'^{-1})$ for all $h, h'\in SO(3)$ \cite{FrankelBook}. We have evaluted $J_{\rm eff}(\beta)$ to high order and equating $J_{\mathrm{eff}}(\beta_{\rm chi}) = \beta_{c, \rm 3D Ising}\simeq 0.2215$ converges and gives $\beta_{\rm chi} = 1.2064$ in excellent agreement with our Monte-Carlo value. 

Similarly, we can compute effective mean-field value of $\overline{J}_{\mathrm{eff}}(\beta)$ with the Ising variables $\sigma_{\rm MF} \sim \sigma_i\sigma_j$ from
\begin{align} \label{eq:JeffMF_SO3}
\overline{J}_{\mathrm{eff, MF}}(\beta) &= 
\corr{\mathrm{Tr}[W_{ij}]}(\beta) \\ 
&\hspace*{-1.5cm}= \frac{1}{Z_{\rm MF}} \sum_{\sigma_{\rm MF} = \pm 1} \prod_{\corr{ij}} \int_{SO(3)} \mathrm{d} W_{ij}~\mathrm{Tr}[W_{ij}] e^{\beta \mathrm{Tr}[W_{ij}] \sigma_{\rm MF}} \nonumber\\
&\hspace*{-1.5cm}=\frac{\big[\beta  \sinh (\beta )+\cosh (\beta )\big] \, _0F_1\left(2;\beta ^2\right)-\cosh
   (\beta ) I_0(2 \beta )}{\cosh (\beta ) I_0(2 \beta )-\sinh (\beta ) I_1(2 \beta )}, \nonumber
\end{align}
where $_0F_1(a;x)$ is the regularized confluent hypergeometric function. The correlator $\corr{\mathrm{Tr} [W_{ij}]}$ leading to $\overline{J}_{\rm{eff, MF}}$ is again independent of $i,j$ since the link variables $W_{ij}$ are free in the mean-field approximation for the Ising variables. The theoretical line of $J_{\rm{eff}, MF}(\beta)$ for $SO(3)$ is shown in Fig. \ref{fig:JeffMF_SO3} for comparison with the Monte-Carlo results. We see, as expected, that the low and high temperature correlations of $SO(3)$ agree with the MF result and this in fact applies to all point groups. For $SO(3)$ and $T, I, O$ realizing the vestigial chiral phase, closer to the transition temperature $\beta_{\rm chi}$, the fluctuations of the Ising fields reinforce the orientational short-range order significantly from the mean-field result. What is surprising is that the fluctuations continue to be identical to $G=SO(3)$ for $T, O$ and $I$, up until to the regime of the nematic transition.

\end{appendix}

\bibliography{chiralpaper}

\end{document}